# Strain relaxation during the layer by layer growth of cubic CdSe onto ZnSe

O. de Melo,[a] C. Vargas-Hernández,[b] I. Hernández-Calderón
*Physics Department, CINVESTAV-IPN, Ave. IPN 2508, 07000 Mexico*

A detailed reflection high-energy electron diffraction analysis shows relevant features of the lattice parameter relaxation of CdSe thin films grown in a layer-by-layer mode onto ZnSe. *In situ* investigations of different azimuths show a clear lattice parameter oscillation in the [110] azimuth. The lattice parameter has a minimum value (similar to that of ZnSe) during Se exposure steps, and a higher and increasing lattice parameter during Cd exposure steps. The behavior is ascribed to the formation of CdSe islands during Cd exposure steps. The cumulative effect in CdSe exposure steps is considered to be a consequence of a decrease in the island size with the number of cycles. Actual plastic deformation does occur after 5 ML. © *2003 American Institute of Physics.*
[DOI: 10.1063/1.1534941]

Lattice relaxation in strained thin film semiconductors has became a very important point recently. Following the standard accepted criterion, it can be considered that the strained overlayer suffers a tetragonal distortion with the in-plane lattice parameter being equal to that of the substrate and the transverse lattice parameter being reduced (enlarged) for in-plane tensile (compressive) strain, according to the elasticity theory. This pseudomorphic growth continues until the stress is relaxed through misfit dislocation formation or (for largely mismatched systems) three-dimensional (3D) island formation. However, evidence has been reported of a much more complicated relaxation process[1] in which the lattice parameter has been observed to oscillate during molecular beam epitaxy (MBE) growth of $In_xGa_{1-x}As$ onto GaAs substrates. The effect has been also observed in some metallic thin films[2,3] and in the II–VI semiconductor ZnTe grown onto CdTe,[4] always using the MBE technique. It has been ascribed to the relaxation of the strain in the edges of the growing islands. A strong anisotropy in the relaxation process has been noticed concurrently with the oscillation of the lattice parameter.[3,4] The geometry of the islands or the stress induced by surface contamination or reconstruction has been invoked to explain this anisotropy. Anisotropic lattice parameter oscillations have been also observed in metallic and semiconducting[5] homoepitaxial structures. In this last reference both MBE and atomic layer epitaxy (ALE) were used for growing CdTe homoepitaxial films.

In this letter we report the occurrence of lattice parameter oscillations in the CdSe/ZnSe system. To explore this highly strained system we used ALE instead of MBE used in the earlier studies of heteroepitaxial films. In our growth conditions, the ALE regime was observed (it was verified in previous CdSe growth experiments) to be self-regulated at 0.5 ML per cycle. For this, we could separate the influence of the roughness induced by fractional surface coverages and that due to the different morphologies of Cd and Se terminated surfaces, as will be explained later.

The CdSe/ZnSe structures were grown in a RIBER 32P MBE system onto semi-insulating GaAs (001) substrates at 260 °C. As received episubstrates were introduced in the MBE chamber and then thermally deoxidized. Zn/Se and Cd/Se beam pressure ratios were 1/3 and the growth rate for MBE ZnSe was about 0.7 $\mu$m/h. Five CdSe quantum wells of 4 MLs were grown by ALE separated by 20-nm-thick ZnSe barriers onto a 500 nm ZnSe buffer layer. To improve two-dimensional (2D) growth, before the growth of the buffer layer and before the growth of every CdSe quantum well (QW), 15 cycles of ALE ZnSe were performed. Reflection high-energy electron diffraction (RHEED) patterns were recorded in the [110], [100], and [1$\bar{1}$0] azimuths for different wells. The exposure and dead times were 15 and 5 s, respectively. It was observed that the quality of the ZnSe RHEED patterns was rapidly recuperated after the deposition of the ZnSe barriers over the CdSe wells and had a very similar appearance to that of the ZnSe buffer layer. To verify that the same initial conditions were reproduced after every ZnSe barrier, the [110] azimuth was investigated in the first and the fifth QWs and the same features were observed in both cases. RHEED patterns images were digitized and the distance between streaks could be measured with relatively good accuracy. Cd exposed surfaces presented a typical $C(2\times 2)$ surface reconstruction characterized by half order reconstruction in the [100] azimuth. Se exposed surfaces presented a $(2\times 1)$ reconstruction characterized by half order streaks in the [110] azimuth. This half order streaks can be observed in the intensity profiles of Figs. 2 and 3 thereafter. Figure 1(a) represents the intensity profile of the RHEED pattern in the [110] azimuth of the ALE growth of the CdSe wells taken during Cd exposure. The interstreak distance ($t$) decreases with increasing cycle number (arrows are a guide to the eye). The distance between the central and the first streak was reduced from 36.5 to 34.5 pixels. This represents a decrease of 5.5%, which can be compared with the CdSe/ZnSe lattice parameter ratio (106.7%). Considering the inverse dependence between $t$ and the lattice parameter, it is possible to calibrate $t$ to obtain the dependence of the in-plane lattice parameter corresponding to this [110] azimuth with the number of cycles. Figure 1(b) represents such dependence during the Cd exposure steps of the ALE cycles.

[a]Present address: Physics Faculty, University of Havana; electronic mail: omelo@ff.oc.uh.cu
[b]Present address: Universidad Nacional de Colombia, Sede Manizales, Departamento de Fisica y Quimica, AA. 127 Manizales, Columbia.

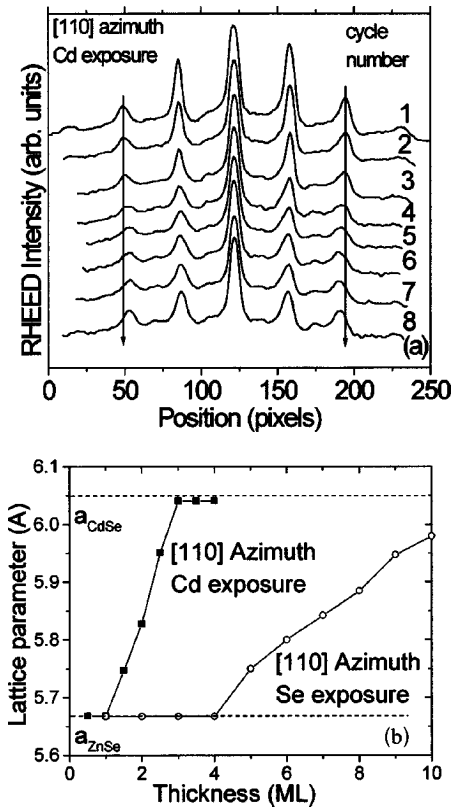

FIG. 1. (a) RHEED intensity profiles for Cd exposed surfaces in the [110] azimuth. (b) Lattice parameter variation as a function of the number of cycles calculated from the inter-streak distances in the [110] azimuth for Cd (solid squares) and Se (open circles) exposures (2 cycles=1 ML).

We did not observe any 2D to 3D growth mode transition: the RHEED pattern keeps its streaky character during the whole growth. Another important characteristic observed in Fig. 1 is that the modification of $t$ is guided through an increase in the width of the streaks towards the center of the diffraction pattern. In the case of a tetragonal distortion, a similar behavior of $t$ would be expected in the other azimuths on Cd and Se exposures. However, our measurements in the [1$\bar{1}$0] azimuth revealed no modification in $t$. This is shown in Fig. 2(a) where the first and eighth cycles intensity profiles (again during Cd exposure) are shown. This result can be explained assuming that the relaxation process leads to a distortion of the averaged reciprocal lattice, i.e., distance between rods are decreased only when observed from the [110] azimuth. To verify this hypothesis we studied the evolution of the RHEED pattern in the [100] azimuth [Fig. 2(b)]. In this case (also) we observed a decrease in $t$ with the number of cycles; this decrease, however, was lower than that observed in the [110] azimuth. This average distance was reduced only from 51 to 49.8 pixels (4.4%).

The profiles for the first and eighth cycles during Se exposure in the studied azimuths are shown in Fig. 3. In the [110] azimuth profile, less intense peaks coming from the (2×1) reconstructed surface are observed at half way between bulk lattice ones. We do not observe any appreciable change in $t$ for any azimuth. Moreover, the width of the streaks keeps approximately constant, indicating a homogeneous lattice parameter. These observations and those of Cd terminated surfaces, indicate an oscillatory behavior of the lattice parameter in alternating Cd–Se exposures. This is the same effect previously observed in other heterostructures.[1–4] In the present case, however, the period of the oscillation does not coincide with the deposition time of 1 ML but with that of 0.5 ML according the autoregulated (at 0.5 ML per cycle) ALE regime used in our experiments. For this, we assume that the oscillation of the interatomic distance of surface atoms (observed in the [110] azimuth) is the consequence of a very different surface morphology in Cd and Se terminated surfaces. In previous MBE experiments (cited above) the period of the oscillation was equal to the time of deposition of one monolayer and the surface morphology was rather guided by the dynamics of layer-by-layer growth and the coverage of the surface. Following the argument of relaxation at the edges of the islands we can infer that Cd surfaces are rougher and present a higher island density; during Se deposition a smoother surface is recuperated and the ZnSe lattice parameter predominates. According the atomic model for the autoregulated growth rate at 0.5 ML/cycle in CdTe,[6] the surface roughness variation would have a period of 2 ALE cycles (1 ML). In that article, observed oscillations of the specular spot intensity with a period of 1 ALE cycle (with lower intensity during Cd exposure steps) are due to the chemical signal associated with different reconstructions during anion or cation exposures. A two cycle period oscillation was detected as a superimposed modulation to this chemical signal. In Ref. 5, oscillations of the lattice parameter are observed in homoepitaxial ALE CdTe. There, a relative large modification on the average lattice parameter is observed only on Cd exposure, while a much smaller (and with opposite sign) modification is present during Te exposure. The origin of the lattice oscillation was explained as due to the stress induced by the different surface reconstructions. In our case, the strain is largely dominated by the mismatch in the CdSe/ZnSe heterostructure. Our consideration of a different surface morphology in the same ALE cycle presupposes that a rearrangement of the surface atoms occurs in every exposure to Cd or Se. This process would be guided by a large stress (added to the mismatch induced one) related with the $C(2\times 2)$ reconstruction on Cd exposure. On Se exposure, the $(2\times 1)$ reconstruction would allow less stress and the surface is smoothed. The cumulative effect (increasing of the lattice parameter up to the value of CdSe) in successive Cd exposures can be explained as due to a decrease in the size (and an increase in the density) of the CdSe islands with increasing cycle number. For larger islands, as expected for surfaces grown at the initial cycles, the effect of the edges of the islands (with larger lattice parameter due to the distortion) is negligible. In this case, the averaged lattice parameter calculated from the maxima of the intensity in the profiles, is practically equal to that of the ZnSe as corresponding to a pseudomorphic growth. For larger thickness of the CdSe films, the "weight" of the lattice parameter of the island edges on the averaged lattice parameter is enhanced, leading to an increase in the width of the streaks toward the center of the diffraction pattern (as observed in Fig. 1). Finally, a shift of the maximum of the intensity of the streaks up to reach the $t$ corresponding to almost pure CdSe (when the lattice parameter of the edges of the islands predominates) is observed. To check for the real onset of dislocation formation, we performed experiments with larger number of cycles. We observed that complete relaxation is a very gradual process. For example, modification in the interstreak distance in the [110] azimuth during Se exposure starts after 10 ALE cycles as shown in Fig. 1(b).

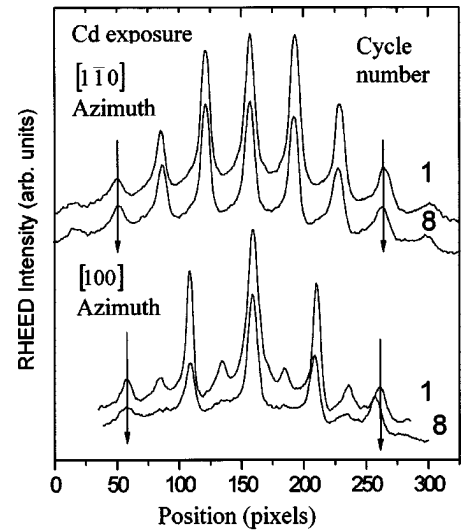

FIG. 2. RHEED intensity profiles for Cd exposed surfaces in the [100] and the [1$\bar{1}$0] azimuths. First and eighth cycles are shown for every azimuth.

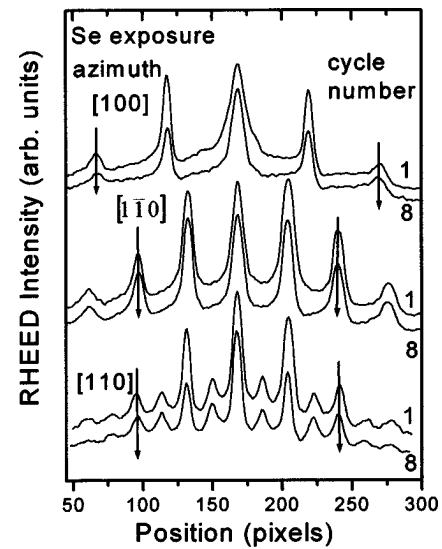

FIG. 3. RHEED intensity profiles for Se exposed surfaces in the [100], [1$\bar{1}$0], and [100] azimuths. First and eighth cycles are shown for every azimuth.

There is still a questionable point in our earlier description of different morphologies on Cd and Se exposures. A completely covered $C(2\times 2)$ reconstructed surface, has only half of the atoms with respect to the bulk normal structure. For this, if the surface is completely covered on Cd exposure, the growth rate will be $\frac{1}{2}$ ML per cycle (1 ML is intended to be a Cd layer plus a Se layer) as expected according the atomic model of Ref. 6 But if there are islands during Cd exposure, as our observations seem to indicate, only a fraction of the surface would be covered with $C(2\times 2)$ arranged Cd atoms. As a consequence, the "area averaged" coverage of Cd will be smaller than $\frac{1}{2}$ ML. This means that the formation of $\frac{1}{2}$-ML-thick islands on Cd exposure is in disagreement with the observed $\frac{1}{2}$ ML per cycle growth rate.

A plausible explanation to account for this disagreement is that the arrangement of the islands is of that kind that the not covered area is negligible or that the actual growth rate in the first stages of the growth is slightly lower than 0.5 ML.

Anisotropy in the relaxation for different azimuths was observed previously by Eymery et al.[4] in ZnTe/CdTe heterostructures grown by MBE. Contrary to our work, they observed the strongest oscillations in the [1$\bar{1}$0] azimuth and not in the [110] one. It was considered to be due in part to the different size of the island in the [110] and [1$\bar{1}$0] directions. Both structures are formed by II–VI semiconductors, have the same zinc blende structure, and have a common anion. Then, a similar behavior of the anisotropy could be expected. Probably, a more important difference between these structures is the sign of the mismatch: ZnTe has a smaller lattice parameter than CdTe while CdSe has a larger lattice parameter than ZnSe. In the former case, atoms at the edge of the islands move toward the center of the island when relaxing the strain; the opposite occurs in the present case. In order to elucidate this point it would be interesting to study the behavior of the relaxation of CdTe onto ZnTe. This would allow comparing the behavior of both systems having the same mismatch sign.

In summary, an oscillatory behavior of the surface lattice parameter was observed during ALE growth of CdSe on ZnSe. The observed period of the oscillation was 1 ALE cycle (or 0.5 ML) in contrast with previous observations in other heterostructures grown by MBE where the period was 1 ML. This peculiarity of the ALE regime, was also noticed in earlier experiments on homoepitaxial ALE CdTe[5] where lattice parameter oscillations with 1 ALE cycle (0.5 ML) period were observed. We conclude that these oscillations are due to different morphologies of Cd and Se exposed surfaces. Cd exposed surfaces present an island-like morphology that leads to an increase in the averaged lattice parameter with the number of cycles. On Se exposure, the smoothness of the surface is recuperated. These observations suggest that the relaxation process of CdSe/ZnSe structure is much more complicated than usually supposed: the observed increase in the lattice parameter at around 2 ML is only a consequence of elastic deformation at the edges of the islands in the Cd exposed surfaces. Actual plastic deformation through misfit dislocations for CdSe grown onto ZnSe (at least in the ALE regime) does occur after 5 ML coverage. These observations would be meaningful when determining critical thickness using RHEED technique.

This work was partially supported by CONACyT-Mexico. The authors appreciate the technical assistance of Z. Rivera, A. Guillén, and H. Silva.